\documentclass[aps,showpacs,showkeys,floatfix,onecolumn]{revtex4}
\usepackage{amsmath,amssymb} 
\usepackage{graphicx} 
\usepackage{dcolumn} 
\newcommand{\field}[1]{\mathbb{#1}}
\newcommand{\R}{\field{R}}
\begin{document}
\title{Critical line of the $\Phi^4$ theory on a simple cubic lattice in the local potential  approximation.}
\date{\today}
\author{\firstname{Jean-Michel} \surname{Caillol}}
\email{Jean-Michel.Caillol@th.u-psud.fr}
\affiliation{Univ. Paris-Sud, Laboratoire LPT, UMR 8627, Orsay, F-91405, France}
\affiliation{CNRS, Orsay, F-91405, France}
\begin{abstract}
We establish the critical line of the  one-component  $\Phi^4$ (or Landau-Ginzburg) model   on the simple  cubic lattice in three dimensions. 
Our study is performed in the framework
of the non-perturbative renormalization group in the local potential approximation. Soft as well as  ultra-sharp infra-red regulators
are both considered.  While the latter gives poor results,  the critical line given by the soft cut-off compares well with
the  Monte Carlo simulations  data of  Hasenbusch \mbox{(\textit{J. Phys. A : Math. Gen.}  \textbf{32}  (1999) 4851)} 
with a relative error of, at worst,  $\sim 3. 10^{-3}$ on published points (critical parameters) of this line.

\end{abstract} 
\maketitle
\section{\label{sec:intro}Introduction}
The non perturbative renormalization group (NPRG) approach initiated by Wetterich \textit{et al.} 
\cite{Wetterich1,Wetterich2,Delamotte} has proved its ability to describe both universal and non universal quantities 
for various models of statistical and condensed matter physics near criticality. It has been extended and adapted to lattice models  
recently \cite{Dupuis-S} and applied successfully to the three-dimensional (3D) Ising, XY, and Heisenberg model \cite{Dupuis-M}. 
Here we apply the lattice NPRG to the one-component $\Phi^4$ model in 3D on a simple cubic lattice. 

We work in the framework of the local potential approximation (LPA) and consider both a sharp and a smooth infra-red cut-off (or
regulator). As in  Refs.~\cite{Bonanno,Caillol1} the flow equations are integrated out for the so-called threshold functions \cite{Wetterich2}
 rather than for the potential. The resulting flow equations turn out to be  quasi-linear parabolic
partial differential equations (PDE) for which  several stable and unconditionally convergent numerical algorithms have been developed 
by mathematicians~\cite{Ames}.   We made use of the algorithm of Douglas-Jones~\cite{Ames,Douglas}  to solve the NPRG flow equations both above
and below the critical temperature; this yields an easy and precise determination of the critical point.
The critical line of the model is obtained for a large range of  parameters and compared  with the Monte Carlo (MC) 
data of Ref.~\cite{Hasenbusch}.

Our paper is organized as follows :
In Sec.~\ref{secI} we give a short review of lattice field theory and of recent advances
in the application  of  the version of Wetterich~\cite{Wetterich1,Wetterich2}  of  the NPRG
to lattice systems \cite{Dupuis-M,Dupuis-S}.
Then, in Sec.~\ref{LPA_Sec},  we detail the LPA approach
which constitutes the simplest non-perturbative approximation to solve the flow equations. Two convenient regulators are introduced
and the corresponding flow equations are derived. Mathematically, the flow equations for the potential are non-linear parabolic
PDE for which the correct initial conditions need a thorough discussion. A numerical solution of these equations requires a change of
variables explained in Sec.~\ref{Change}. After this transformation,  the emergent equations turn out to be  quasi-linear parabolic PDE
and can thus be solved easily on a computer making use of powerful algorithms.
Numerical results for the critical line of the  one-component  $\Phi^4$ model   on the simple  cubic lattice in three dimensions
are reported in Sec.~\ref{Num} and compared  with the Monte Carlo (MC) 
data of Ref.~\cite{Hasenbusch}.  The sharp cut-off regulator leads to poor results while the soft cut-off one reproduces the MC data
with a relative precision of a few $10^{-3}$. We conclude in Sec.~\ref{Conclu}.
\section{\label{secI}Prolegomena}
\subsection{\label{def}Lattice  $\Phi^4$ model}
We consider a lattice field theory defined on a D-dimensional (hyper)cubic lattice. The action
is given by \cite{Montvay}

\begin{equation}
\label{S1}
 \mathcal{S} \left[ \varphi \right] = \frac{1}{N a^D}  \sum_{\left\lbrace \mathbf{q}\right\rbrace } \varphi_{-\mathbf{q}} 
                                           \epsilon_0 \left(  \mathbf{q}   \right)  \varphi_{\mathbf{q}}        + a^D
\sum_{\left\lbrace  \mathbf{r} \right\rbrace} U_0(\varphi_{\mathbf{r}} ) \; ,
\end{equation}
where $a$ is the lattice constant, and $\left\lbrace  \mathbf{r} \right\rbrace$ denotes the $N$ sites of the lattice.
For simplicity we limit ourselves
to a one-component real field $\varphi_{\mathbf{r}}$ and a simple cubic (SC) lattice, $ \varphi_{\mathbf{q}}=a^D\sum_
{\left\lbrace  \mathbf{r} \right\rbrace}
e^{- i \mathbf{r}\mathbf{q}}  \varphi_{\mathbf{r}}$ is the Fourier transform of the field and the $N$ momentum
 $\left\lbrace \mathbf{q}\right\rbrace $ are
restricted to the  first Brillouin zone $]-\pi/a, \pi/a ]^{D}$.  In the thermodynamic limit ($a$ fixed, $N \to \infty$),
\begin{equation}
 \frac{1}{N a^D}  \sum_{\left\lbrace \mathbf{q}\right\rbrace }  \longrightarrow
\int_{-\pi/a}^{\pi/a} \dfrac{d q_1}{2 \pi}  \ldots \int_{-\pi/a}^{\pi/a} \dfrac{d q_D}{2 \pi} \equiv \int_{ \mathbf{q} } \; .
\end{equation}
The local potential $U_0$ is defined such that $\epsilon_0 ( \mathbf{q}=0) = 0$ and henceforth we adopt
the Landau-Ginzburg polynomial form $U_0(\varphi) = (\mathrm{r}/2) \;  \varphi^2 + (\mathrm{g}/4!) \; \varphi^4$. The spectrum 
$\epsilon_0 \left(  \mathbf{q}   \right)$ accounts for next-neighbor interactions; for a SC lattice it reads as
\begin{equation}
 \epsilon_0 \left(  \mathbf{q}   \right) = \frac{2 \xi }{a^2} \; \sum_{\mu=1}^{D}\left( 1-\cos\left(q_{\mu}a \right) \right)  \; ,
\end{equation}
 where the dimensionless parameter  $\xi $ rules the amplitude of the spectrum. 
 $\epsilon_0 \left(  \mathbf{q}   \right) \sim \xi \mathbf{q}^2$
for $ \mathbf{q} \to 0$ and $\max_{ \mathbf{q}} \epsilon_0 \left(  \mathbf{q}\right) = \epsilon_0^{\max}=(4 D \xi )/a^2$.
Note that the dimension of the field is $[\varphi] = D/2 -1$ and that, 
in the thermodynamic limit,  the physics of the model  depends only upon the two dimensionless parameters
 $\overline{ \mathrm{r}}=  \mathrm{r} a^2/ \xi $
and $\overline{\mathrm{g}}= \mathrm{g}a^{(4-D)}/ \xi^2$ 

Another way of writing the action \eqref{S1}, which is useful for numerical investigations, is \cite{Montvay}
\begin{equation}
 \label{S2}
 \mathcal{S} \left[ \psi \right] = \sum_{\left\lbrace \mathbf{n} \right\rbrace }
                                             \left[ -2 \kappa \sum_{\mu=1}^{D} \psi_{\mathbf{n}} \psi_{\mathbf{n}+\mathbf{e}_{\mu} }
                                                    + \psi_{\mathbf{n}}^2 w
                                                    + \lambda \left(  \psi_{\mathbf{n}}^2 -1   \right) ^{2} -\lambda
\right]   \; ,
\end{equation}
where the $D$ unit vectors $\mathbf{e}_{\mu}$ constitute an orthogonal basis set for $\R^D$.
The field  $\psi$ and the parameters $(\kappa, \lambda)$ are all dimensionless awnd are related to the bare field $\varphi$
and dimensionless parameters ($\overline{ \mathrm{r}}$,  $\overline{ \mathrm{g}}$) through the relations
\begin{subequations}
\label{toto}
\begin{align}
  \psi_{\mathbf{n}} &= \sqrt{\frac{\xi}{2 \kappa}}\; a^{D/2 -1} \; \varphi_{\mathbf{r}}  \; \; \mathrm{ with } \; \; \mathbf{r} = a \mathbf{n} \; , \\
\overline{ \mathrm{r}} &=  \dfrac{1 - 2 \lambda}{\kappa} -2 D    \; ,    \\
\overline{ \mathrm{g}} &= \dfrac{6 \lambda}{\kappa^2} \; .
\end{align}
\end{subequations}

To close this section let us recall some elementary definitions and results concerning  lattice field theory  \cite{Montvay}.
The physics (thermodynamic and correlation functions) of the model is coded in the partition function 
\begin{equation}
\label{Z}
 Z\left[  h \right] = \int \mathcal{D} \varphi \exp\left( -S\left[  \varphi \right] + \left(h \vert \varphi  \right)  \right)  \; ,
\end{equation}
where the dimensionless measure is given by
\begin{equation}
 \mathcal{D} \varphi = \prod_{\mathbf{r}}\; \left[a^{D/2 -1} \; d\varphi_{\mathbf{r}} \right]  \; ,
\end{equation}
 $h$ is an external lattice field and the scalar product in~\eqref{Z} is defined as 
\begin{equation}
 \left(h \vert \varphi  \right)  = a^D \sum_{\mathbf{r}} h_{\mathbf{r}} \varphi_{\mathbf{r}} \; .
\end{equation}
The order parameter is given by 
\begin{equation}
 \phi_{\mathbf{r}} =  \left\langle \varphi_{\mathbf{r}} \right\rangle =     \dfrac{1}{a^D}    \frac{ \partial W\left[  h \right]}{\partial h_{\mathbf{r}}  } \; ,
\end{equation}
where the Helmholtz free energy $W\left[  h \right]= \ln Z\left[  h \right]$ which is a convex function of the $N$
 variables $\left\lbrace h_{\mathbf{r}}\right\rbrace $ is the
generator of the connected correlation functions $G^{(n)}(\mathbf{r}_1 \ldots \mathbf{r}_n) = a^{-nD} 
( \partial / \partial h_{\mathbf{r}_1} )   \ldots ( \partial / \partial h_{\mathbf{r}_n})  W\left[  h \right] $. The Legendre transform of $W\left[  h \right]$,
\textit{i. e. } the Gibbs free energy, will be denoted  unusually by $\overline{\Gamma}\left[  \phi \right]$ and reads as
\begin{equation}
 \overline{\Gamma}\left[  \phi \right] = \left( h\vert \phi \right) - W\left[  h \right]  \; .
\end{equation}
 $ \overline{\Gamma}\left[  \phi \right] $ is a convex function of the $N$  field variables  $\left\lbrace \phi_{\mathbf{r}}  \right\rbrace $ and 
 the generator of the vertex functions $\overline{\Gamma} ^{(n)}(\mathbf{r}_1 \ldots \mathbf{r}_n) = a^{-nD} 
( \partial / \partial \phi_{\mathbf{r}_1} )   \ldots ( \partial / \partial \phi_{\mathbf{r}_n})   \overline{\Gamma}\left[  \phi \right] $. Finally, as well known,
the matrix  $\overline{\Gamma} ^{(2)}(\mathbf{r}_1 , \mathbf{r}_2)$ is the inverse of matrix $G^{(2)}(\mathbf{r}_1 , \mathbf{r}_2)$. 

\subsection{\label{NPRG}Lattice NPRG}

To implement the lattice NPRG  procedure we follow the suggestion of Dupuis \textit{et al.} in Ref. \cite{Dupuis-S, Dupuis-M} which
extends to the lattice the ideas of Wetterich \cite{Wetterich1,Wetterich2} for the continuous version ($a \to 0$) of the model.  
We thus add to the action~\eqref{S1} the regulator term
\begin{equation}
 \label{regu}
\Delta \mathcal{S}_{k}\left[   \varphi \right]  = \frac{1}{2} \frac{1}{N a^D} \sum_{\left\lbrace \mathbf{q}\right\rbrace }
                                     \varphi_{-\mathbf{q}} R_k\left( \mathbf{q}\right)  \varphi_{\mathbf{q}}   \; .
\end{equation}
where $ R_k\left( \mathbf{q}\right)$ is  positive-definite, has the dimension $[R_k]=2$ and acts as a  $\mathbf{q}$ dependent mass term. 
The regulator $R_k \left( \mathbf{q}\right) $ is chosen in such a way that it acts as an infra-red (IR) cut-off which leaves the high-momentum modes unaffected
and gives a  mass to the low-energy ones.  Roughly $R_k \left( \mathbf{q}\right) \sim 0 $ for $\vert \vert \mathbf{q} \vert \vert> k $
and $R_k \left( \mathbf{q}\right) \sim Z_k k^2 $ for $\vert \vert \mathbf{q} \vert \vert <  k $.
The scale $k$ in momentum space varies from $\Lambda \sim a^{-1}$, some undefined microscopic scale of the  model yet to be precised,
to $k=0$ the macroscopic scale. 
To each scale $k$ corresponds a $k$-system defined by its microscopic action $ \mathcal{S}_{k}\left[   \varphi \right]=
\mathcal{S} \left[   \varphi \right] + \Delta \mathcal{S}_{k}\left[   \varphi \right] $. We denote its partition function by $Z_k\left[ h \right] $, its Gibbs free energy by 
$ \overline{\Gamma}_k\left[  \phi \right]$, \textit{etc}.

For technical reasons that should become clear below, we are rather interested in the so-called average effective action
 $\Gamma_k \left[\phi \right] $ which was introduced by Wetterich~\cite{Wetterich1} and is 
defined as a modified Legendre transform of $W_k\left[ h \right] $ which includes the explicit substraction of
 $\Delta \mathcal{S}_{k}\left[   \phi \right]$ \cite{Wetterich2}, \textit{i. e.}
\begin{equation}
  \Gamma_k \left[\phi \right] = \overline{\Gamma}_k\left[  \phi \right] -\Delta \mathcal{S}_{k}\left[   \phi \right] \; .
\end{equation}
Note that the functional $\Gamma_k \left[\phi \right] $ is not necessarily convex by contrast with $ \overline{\Gamma}\left[  \phi \right]$
which is the true Gibbs free energy of the "k``- system. 
It satisfies the exact flow equation \cite{Wetterich1,Wetterich2,Delamotte,Dupuis-S,Dupuis-M}
\begin{equation}
 \label{flow}
\partial_k \; \Gamma_k\left[ \phi \right] = \frac{1}{2}  \; \sum_{\mathbf{q}} \partial_k R_k\left( \mathbf{q}\right) 
                      \left[  \Gamma_k^{(2)}  + R_k  \right] ^{-1}_{\mathbf{q}, -\mathbf{q} } \; .
\end{equation}
 Note that Eq.~\eqref{flow} is an extremely complicated equation since the vertex function $\Gamma_k^{(2)}(\mathbf{q},-\mathbf{q})$,
 which is the Fourier transform of the second-order functional derivative
of $\Gamma\left[ \phi \right] $ with respect to the classical field $\phi$, depends functionally upon  $\phi$.
For an homogeneous configuration of the field $\phi_{\mathbf{r}}= \phi$ we have, on the one hand, 
 $\Gamma_k\left[ \phi \right]= N a^D U_k(\phi) $ where
the potential $U_k(\phi)$ is a simple function of the field $\phi$ and, on the orther hand, the conservation of momentum 
at each vertex which implies, with the usual 
abusive notation,  $ \Gamma_k^{(2)}(\mathbf{q},- \mathbf{q})= N a^D \Gamma_k^{(2)}(\mathbf{q})$ from which follows :

\begin{subequations}
\label{flow_2}
\begin{align}
\partial_k U_k ( \phi )  &= \frac{1}{2}\frac{1}{N a^D}  \; \sum_{\mathbf{q}} \frac{\partial_k R_k\left( \mathbf{q}\right) }{  \Gamma_k^{(2)}(\mathbf{q})
                    + R_k\left( \mathbf{q}\right)} \; , \\
&=  \frac{1}{2}  \int_{\mathbf{q}} \frac{\partial_k R_k\left( \mathbf{q}\right) }{  \Gamma_k^{(2)}(\mathbf{q})
                    + R_k\left( \mathbf{q}\right)} \; , \label{limit_TH}
\end{align}
\end{subequations}
where the second line~\eqref{limit_TH} is valid  in the thermodynamic limit ($a$ fixed, $N \to \infty$).

We can give a formal solution of~\eqref{flow} as \cite{Wetterich2,Delamotte} 
\begin{equation}
 \label{Gimplicit}
\exp\left( -\Gamma_k\left[ \phi \right]  \right) =
\int \mathcal{D}\phi \; \exp \left( -\mathcal{S}\left[ \varphi \right] + (\varphi - \phi \; \vert \dfrac{\delta\Gamma_k[\phi]}{\delta \phi})
-\frac{1}{2} (\varphi - \phi\; \vert R_k \vert \varphi - \phi\;)
 \right)
\end{equation}
which gives $\Gamma_k\left[ \phi \right] $ implicitly. Eq.~\eqref{Gimplicit} allows us to precise the initial conditions. The initial
value $k= \Lambda$ of the momentum scale $k$ is chosen such that $R_{\Lambda}(\mathbf{q}) \sim \infty$ for all values of $\mathbf{q}$ hence,
since $\exp( -1/2 \; (\chi \vert R_{\Lambda} \vert \chi )) \propto \delta[\chi])$,  where $\delta[\chi]$ is the Dirac functional, 
it follows from~\eqref{Gimplicit} that
$\Gamma_{\Lambda}[ \phi] =\mathcal{S}[ \phi]$.
Physically it means that all fluctuations are frozen and the mean-field theory becomes exact. When the running momentum goes from $k=\Lambda$
to $k=0$ all the modes $\varphi_{\mathbf{q}}$ are integrated out progressively and the effective average action evolves from its microscopic limit
$\Gamma_{\Lambda}[ \phi] = \mathcal{S}[ \phi]$ to its final macroscopic expression $\Gamma_{k=0}[ \phi] =\Gamma[ \phi]$. The choice of initial
momentum $\Lambda$ which depends on the choice of the regulator $R_k$  will be made more explicit in Sec.~\ref{LPA_Sec}.
\section{\label{LPA_Sec}Local potential approximation}

An increasingly popular  way to solve the flow eq.~\eqref{flow} is to make an ansatz on the functional form of the effective average action
$\Gamma_k[\phi]$. In the
lattice  LPA
one neglects the renormalization of the spectrum and assume the local \mbox{ form  \cite{Dupuis-S,Dupuis-M}}
\begin{equation}
 \label{LPA}
\textrm{(LPA ansatz)} \; \; \Gamma_k\left[ \phi \right] =  \frac{1}{N a^D}  \sum_{\left\lbrace \mathbf{q}\right\rbrace } \phi_{-\mathbf{q}} 
                                           \epsilon_0  \left(  \mathbf{q}   \right)  \phi_{\mathbf{q}}        + a^D
\sum_{\left\lbrace  \mathbf{r} \right\rbrace} U_k(\phi_{\mathbf{r}} ) \; .
\end{equation}
For a uniform configuration of the classical field $\phi_{\mathbf{r}} = \phi$ and, in the thermodynamic limit, 
 the flow equation~\eqref{limit_TH} becomes : 
\begin{equation}
\label{flow_LPA}
\partial_k U_k ( \phi ) 
 =  \frac{1}{2}  \int_{\mathbf{q}} \frac{\partial_k R_k\left( \mathbf{q}\right) }{  \epsilon_0(\mathbf{q})
                    + R_k\left( \mathbf{q}\right) + U_{k}^{''}(\phi)} \; ,
\end{equation}
where $U_{k}^{''}(\phi)$ denotes the second-order derivation of $U_k(\phi)$ with respect to the order parameter $\phi$.
Eq.~\eqref{flow_LPA} is a non-linear parabolic PDE. It must be supplemented by an initial condition
(see Sec.~\ref{LITIM} and~\eqref{USCO}). Moreover, 
for a numerical resolution of~\eqref{flow_LPA},  boundary conditions for the potential  $U_k$ or one of its derivatives  
  $U_k^{(n)}$,   must also be specified for some maximum
value of the field $\pm \phi_{\textrm{max}}$ (see Sect.~\ref{Change}). Initial and boundary conditions depend on the 
choice of the regulator $R_k$ and, in this paper,  we will consider two possibilities for $R_k$.
\subsection{\label{LITIM}Litim-Machado-Dupuis (LMD) regulator}

In Ref.~\cite{Dupuis-M} Machado and Dupuis consider
\begin{equation}
\label{LMD}
 R_k(\mathbf{q}) = \left[ \epsilon_k - \epsilon_0(\mathbf{q}) \right] 
        \; \Theta\left[  \epsilon_k - \epsilon_0(\mathbf{q})   \right]  \; ,
\end{equation}
which is adapted from Ref.~\cite{Litim} to the lattice case. In Eq.~\eqref{LMD},  $\epsilon_k=\xi \; k^2$ and
$\Theta(x)$ is the step function. This regulator $R_k(\mathbf{q})$ leaves the high-momentum modes 
($\epsilon_0(\mathbf{q}) > \epsilon_k$) unaffected and ascribes a mass $\epsilon_k$ to the low-energy ones.
The effective spectrum of the k-system is obviously
 $\epsilon_k^{\mathrm{eff.}}(\mathbf{q})=\epsilon_0(\mathbf{q}) + R_k(\mathbf{q}) $. We note that
for $k > k_{\mathrm{max}}$, where $k_{\mathrm{max}}$ is defined by 
$\epsilon_{ k_{\mathrm{max}}}= \epsilon_0^{\mathrm{max}}$, \textit{i. e.} $k_{\mathrm{max}}= \sqrt{4 D/a ^2}$,
the effective spectrum $\epsilon_k^{\mathrm{eff.}}(\mathbf{q})= \epsilon_k$ does not depends on $\mathbf{q}$
and we deal with a  theory of $N$ independent sites. 

With that choice the flow equation~\eqref{flow_LPA} takes then a simple synthetic form
\begin{equation}
\label{flow_LMD_f}
 \partial_{t}U_k = - \frac{\epsilon_k}{\epsilon_k + U_k^{''}} \;\mathcal{N}(\epsilon_k) \; 
\end{equation}
where the RG time "$t$" is defined by $k=\Lambda e^{-t}$, so that $\partial_t = -k \partial_k$, and 
\begin{equation}
 \mathcal{N}(\epsilon_k)  = \int_{\mathbf{q}} \Theta (\epsilon_k - \epsilon_0(\mathbf{q})) \; .
\end{equation}
denotes the (normalized) number of states. It will prove convenient to introduce also the density of states
\begin{equation}
\label{D}
 \mathcal{D}(\epsilon_k)  = \int_{\mathbf{q}} \delta (\epsilon_k - \epsilon_0(\mathbf{q})) \; ,
\end{equation}
so that 
\begin{equation}
  \mathcal{N}(\epsilon)  = \int_{0}^{\epsilon} \; d \epsilon^{'} \;  \mathcal{D}(\epsilon^{'})  \; .
\end{equation}

\begin{figure}[t!]
\includegraphics[angle=0,scale=0.55]{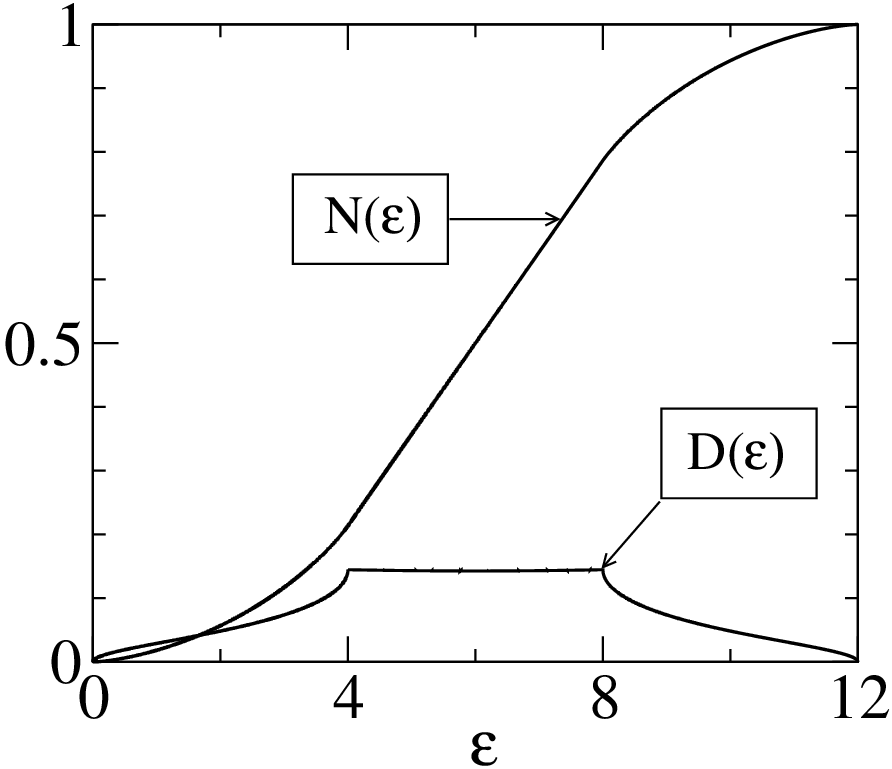}
\caption{
\label{D_N} Density of states $\mathcal{D}(E)$ and number of states $\mathcal{N}(E)$
of the simple 3D cubic lattice
}
\end{figure}
These two functions  $\mathcal{D}(E)$ and $\mathcal{N}(E)$ are obviously related to the lattice Green function 
which, for a SC lattice,  reads \cite{JAP1,JAP2}
\begin{equation}
\label{Green}
 G(\tau) = \frac{1}{\pi^D} \int_0^{\pi}dq_1 \ldots \int_0^{\pi}dq_D  \frac{1}{\tau - \sum_{\mu = 1}^{\mu = D} \cos(q_{\mu})}  \; .
\end{equation}
With the remark that for $\eta \to 0$,  $1/(\tau + i \eta ) = \mathcal{P}(1/\tau)+ i \pi \delta(\tau)$,  the comparison of eqs~\eqref{D} and~\eqref{Green}
reveals that
\begin{equation}
 \mathcal{D}(\epsilon) = \frac{a^{2-D}}{2 \xi } \frac{1}{\pi} \Im{G} (\tau) \; ,
\end{equation}
with $\tau = D -( a^2/2) (\epsilon / \xi)$. Note that the interval of the spectrum
 $0 \leq \epsilon_k \leq \epsilon_0^{\mathrm{max}}$ corresponds
to the interval $-D \leq \tau \leq D$ for the auxiliary variable $\tau$.
In the case of $D=3$ the imaginary part of the  Green function
$G(\tau )$ is given by  \cite{JAP2}

\begin{align}
\label{G_DES}
& 1\leq \tau \leq 3 & \; \; &   \Im{G}(\tau) = \dfrac{1}{\pi^2} \int_0^{cos^{-1}(\tau-2)} dx \;
  K\left(     \dfrac{\left(\chi^2 -1 \right)^{1/2} }{\chi}\right)  \nonumber \\
& 0 \leq \tau \leq 1 &\; \; &  \Im{G}(\tau) = \dfrac{1}{\pi^2} \int_0^{\pi} dx \;
  K\left(     \dfrac{\left(\chi^2 -1 \right)^{1/2} }{\chi}\right)  \; ,
\end{align}
where $\chi= 2/(\tau - \cos(x))$ and $K(y)$ is the complete elliptic integral of the  second kind
\begin{equation}
 K(y) = \int_0^{\pi/2} d\theta \; (1 - y^2 \sin^2(\theta))^{-1/2}  \; .
\end{equation}
One can point out the properties $ \Im{G}(\tau) =\Im{G}(-\tau)  $  for $-3\leq \tau \leq 3$ which implies that
$\mathcal{D}(\epsilon)=\mathcal{D}(\epsilon_0^{\mathrm{max}} - \epsilon) $ and 
$\mathcal{N}(\epsilon_0^{\mathrm{max}} - \epsilon) = 1 -  \mathcal{N}(\epsilon)$ for 
$0\leq \epsilon \leq \epsilon_0^{\mathrm{max}}$.

We made use of the relations~\eqref{G_DES} to evaluate numerically   $\mathcal{D}(E)$ and $\mathcal{N}(E)$, their graphs
are displayed in Fig~\eqref{D_N}. The tiny wiggles in the central part of $\mathcal{D}(\epsilon)$ are actual and could not be avoided,
they  reveal the difficulty to compute this function  with the highest numerical precision. A numerical filter can be used to suppress
the numerical fluctuations in $\mathcal{D}(E)$ and $\mathcal{N}(E)$ before the latter 
is injected in the flow equations~\eqref{flow_LMD_f}.

\subsubsection{The local regime : $ \Lambda \geq k \geq k_{\mathrm{max}}$ }

Clearly one can distinguish 2 different regimes in the flow.
For $\epsilon_k$ larger than the gap of the spectrum,  $\epsilon_k= \xi  k^2 > \epsilon_0^{\mathrm{max}}$, \textit{i. e.}
 $k \geq k_{\mathrm{max}}=\sqrt{4 D/a^2}$, 
we have already pointed out that  the "effective" spectrum of the k-system $\epsilon_k^{\mathrm{eff.}}= \epsilon_k$ 
does not depends on $\mathrm{q}$.
We deal with a local
theory for which the partition function $Z_k\left[ h \right]= \prod_{\mathbf{r}}  z_k(h_{\mathbf{r}})$
is a product of one-site partition functions with

\begin{equation}
\label{z}
 z_k(h) = \int_{-\infty}^{ +\infty} d\varphi \; \exp \left(  -U_0(\varphi)  - \frac{1}{2} \epsilon_k \varphi^2 + h \varphi \right)  \; .
\end{equation}
It follows from this remark that the effective average action has also  a local form
\begin{equation}
 \Gamma_k\left[\phi \right] =  \frac{1}{N a^D}  \sum_{\left\lbrace \mathbf{q}\right\rbrace } \phi_{-\mathbf{q}} 
                                           \xi \left(  \mathbf{q}   \right)  \phi_{\mathbf{q}}  + a^D \sum_{\mathbf{r}} U_{k}^{\mathrm{loc.}} (\phi_{\mathbf{r}}) \; ,
\end{equation}
where the effective average potential $U_k^{\mathrm{loc.}}$ is given implicitly by
\begin{equation}
\label{local_implicit}
 \exp\left(- U_k^{\mathrm{loc.}}\left(\phi \right)  \right) = 
 \int_{-\infty}^{ +\infty} d\varphi \;
\exp\left(
-U_0(\phi) + U^{\mathrm{loc.}\; '}_{k}(\phi) (\varphi - \phi) - \frac{1}{2} \epsilon_k (\varphi - \phi) ^2  
 \right)  \; ,
\end{equation}
as follows from eqs~\eqref{Gimplicit} and \eqref{z}.

\begin{figure}[t!]
\includegraphics[angle=0,scale=0.55]{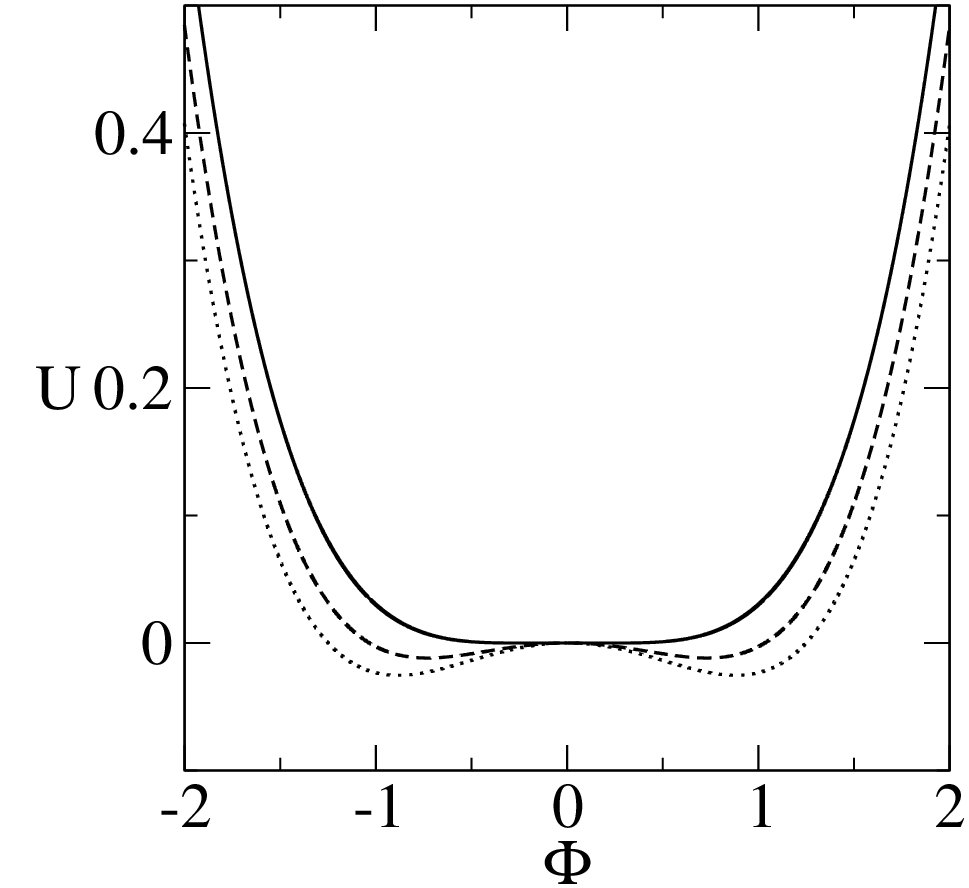}
\caption{
\label{U} Local potential 
of the $\Phi^4$ model on a 3D simple cubic lattice at $g=1$, and $r=-0.13$.
Dotted line : mean field approximation ($\Lambda=5000 a^{-1}$),
Dashed line : local theory at $k=k_{\textrm{max}}$,
Solid line : $U_{k=0}$ in the LPA approximation  with LMD  regulator.
}
\end{figure}

Two remarks are now in order.
Firstly the choice $\Lambda = \infty$ implies $U_\Lambda=U_0$ since we can replace the gaussian 
$\exp(-1/2 \epsilon_{\Lambda} (\varphi - \phi)^2)$
by a delta function $\delta(\varphi - \phi)$ in eq.~\eqref{local_implicit}. Secondly in the range $ \Lambda \geq k  \geq k_{\mathrm{max}}$  the potential $U_k$
satisfies the exact flow equation 
\begin{equation}
\label{zou}
  \partial_{t}U_k = - \frac{\epsilon_k}{\epsilon_k + U_k^{''}}  \; ,
\end{equation}
as shown in the appendix.
Note that for  $ \Lambda \geq k  \geq k_{\mathrm{max}}$  we have $\mathcal{D}(\epsilon_k) = 1$ and thus the LPA flow equation~\eqref{flow_LMD_f}
is exact for a local theory. This point has been checked numerically in Ref.\cite{Dupuis-M} and is proved mathematically in the appendix.
The initial condition of the flow can thus be chosen 
\begin{itemize}
 \item either $ \Lambda = \infty $ and $U_{\Lambda} = U_0$ (Mean field theory) 
 \item or  $\Lambda= k_{\mathrm{max}}= \sqrt{4 D / a^2}$ and  $ U_{\Lambda} \equiv U^{\mathrm{loc.}}_{k_{\mathrm{max}}} $ \; .
\end{itemize}
In the second case the local partition function~\eqref{z} must be computed numerically. It turns out that integrating out the flow
equation~\eqref{zou} with the  mean field initial condition (numerically with a  large value of  $\Lambda$, \textit{e. g. } $\Lambda \sim 5000 a^{-1}$)
 gives more accurate results
than the direct calculation and manipulations of $z(h)$ which involve too large arguments in the exponentials.

We exemplify this discussion in Fig.~\ref{U} where we display, for the $\Phi^4$ model at $r=-0.13$ and 
$g=1$ (\textit{i. e.} in the ordered phase),
$U_{\Lambda}(\phi)=U_0(\phi)$ (mean field approximation), $U_{k_{\mathrm{max}}}(\phi)$ (local theory) and the renormalized potential
$U_k(\phi)$ at $k=0$ obtained after integration of the flow equation~\eqref{flow_LMD_f} 
(the figure also illustrates the passage to convexity : note the flat part of $U_{k=0}(\phi)$  in the range $(-\phi_0, \phi_0$)).

\subsubsection{The non-trivial regime : $ k_{\mathrm{max}} \geq k \geq 0$ }
In this range of $k$ the flow is non-trivial and must be integrated out numerically.  For convenience
we rewrite~\eqref{flow_LMD_f} as
\begin{equation}
\label{flow_LMD_final}
 \partial_t U_k= - \mathcal{N}(\epsilon_k) \mathcal{L}^{LMD}(\omega_k) \; ,
\end{equation}
where $\omega_k (\phi) \equiv U_k^{''}(\phi) /\epsilon_k$ is a dimensionless renormalized susceptibility and 
\begin{equation}
 \mathcal{L}^{LMD}(x) = \frac{1}{1 + x} \; ,
\end{equation}
is the threshold function \cite{Wetterich2} which takes a very simple expression with the LMD
regulator.

Note that in the limit $k \to 0 $
\begin{align}
 \mathcal{N}(\epsilon) &  \sim \int_{\mathbf{q}} \Theta(k^2 -\mathbf{q}^2 ) \nonumber \\
                                   &  \sim \frac{1}{(2 \pi)^D} \int_0^k dq \; q^{D-1} S_D \nonumber \\
                                   &  \sim \frac{4 v_D}{D} k^D \; ,
\end{align}
where $S_{D}=2 \pi^{D/2} / \Gamma( D/2)$ is the surface of the $D$-dimensional hypersphere and
 $v_D^{-1} =2^{D+1} \pi^{D/2} \Gamma(D/2)$ a numerical factor. Therefore in this limit the flow eq.~\eqref{flow_LMD_f}
reduces to

\begin{equation}
\label{flow_asympt}
 \partial_t U_k = \dfrac{-4 v_D}{D} \; k^D  \mathcal{L}^{LMD}(\omega_k) \; ,
\end{equation}
which is identical to the LPA flow equation with Litim's regulator for the continuous (off-lattice)  theory \cite{Delamotte,Litim,Caillol1}.
 We conclude that
the lattice (\textit{cf.}~eq.~\eqref{flow_LMD_f})  and off-lattice (\textit{cf.}~eq.~\eqref{flow_asympt}) flow equations 
have the same asymptotic properties for $k \to 0$. 
The adimensionned versions of these equations
share thus the same fixed points and the same critical exponents. All these quantities have been computed with the highest numerical 
accuracy in $D=3$ dimensions, see \textit{e.g.} Refs~\cite{Caillol1,BBG1,BBG2}.  Recall that in the LPA, Fisher's exponent 
$\eta=0$ in all dimensions of space (no renormalization of the spectrum) and that the other critical exponents are non-trivial and differ 
from the exact ones by a few per cents in $D=3$.

\subsection{\label{USCO}Ultra-sharp regulator}
The ultra sharp cut-off (USCO) was first introduced by Wegner-Houghton \cite{WH} and considered by many authors in different 
NPRG studies  of the continuous (off-lattice) $\Phi^4$ model \cite{Ni1,Ni2,Hasen}.
In its lattice version, it also yields simple flow-equations. In this case, adapting the definition of Wetterich~\cite{Wetterich2},
 the regulator is defined as
\begin{equation}
\label{USCO_R}
 R_k(\mathbf{q}) = Z  \epsilon_k \Theta\left[  \epsilon_k - \epsilon_0(\mathbf{q})   \right]  \; ,
\end{equation}
where the constant $Z$ is ultimately set to $+ \infty$ \cite{Wetterich2}. In order to deal with the discontinuity of  $R_k(\mathbf{q}) $,
we first introduce a smoothened version $\Theta_{\epsilon}(x) $ of the step function which varies mildly from 0 to 1 in the interval
$(-\epsilon/2, + \epsilon/2)$.  Let $\delta_{\epsilon}(x) = \partial_x \Theta_{\epsilon}(x)$ be the smoothened version of the Dirac
generalized function. Then the flow equation~\eqref{flow_LPA} takes  the form :

\begin{equation}
\label{blito}
 \partial_k U_k (\phi) = \frac{1}{2} \int_{\mathbf{q}}  \frac{Z \epsilon_k \partial_k \epsilon_k\; \delta_{\epsilon} \left( \epsilon_k - \epsilon_0(\mathbf{q})  \right)}
                         {\epsilon_0(\mathbf{q}) + Z \;  \epsilon_k \;\Theta_{\epsilon}\left(  \epsilon_k - \epsilon_0(\mathbf{q})     \right) + U_k^{''}(\phi) } +
                           \frac{1}{2} \int_{\mathbf{q}}  \frac{ Z  \partial_k \epsilon_k \; \Theta_{\epsilon}( \epsilon_k - \epsilon_0(\mathbf{q}              )  )}
                            {\epsilon_0(\mathbf{q})  + Z \epsilon_k    + U_k^{''}(\phi) }         
\end{equation}

The limit $\epsilon \to 0$ for the ill-defined  first term in the r.h.s. of~\eqref{blito} can be taken by making use of an extension 
of a  lemma due  Morris \cite{Morris} which  states that, for $\epsilon \to 0$ 
\begin{equation}
\label{lemme}
\lim_{\epsilon \to 0}   \delta_{\epsilon}(\epsilon_k -  \epsilon_0(\mathbf{q})  ) f( \Theta_{\epsilon}(\epsilon_k -  \epsilon_0(\mathbf{q}), \mathbf{q})  ) =
                \delta(\epsilon_k -  \epsilon_0(\mathbf{q})  ) \; \int_{0}^{1}dt \; f(t,\mathbf{q}) \;,
\end{equation}
provided that the function $f( \Theta_{\epsilon}(q,k),k)$ is continuous at $k=q$ in the limit $\epsilon \to 0$,
which is the case here. This yields
\begin{equation}
  \partial_k U_k (\phi) = \frac{1}{2} (\partial_k \epsilon_k)  \mathcal{D}(\epsilon_k)   \ln\left[
\frac{\epsilon_k + U_k^{''}(\phi)+Z \epsilon_k}{\epsilon_k + U_k^{''}(\phi) }
 \right]  
+\frac{1}{2} (\partial_k \ln \epsilon_k)  \; \mathcal{N}(\epsilon_k)  + \mathcal{O}(Z^{-1})
 \; .
\end{equation}
The last step is to take the limit $Z \to \infty$ of the above eq. with the final result
\begin{equation}
\label{flow_USCO}
  \partial_t U_k= - \epsilon_k \; \mathcal{D}(\epsilon_k) \mathcal{L}^{USCO}(\omega_k) \; ,
\end{equation}
where $\omega_k (\phi) \equiv U_k^{''}(\phi)/\epsilon_k$ and  the USCO threshold function reads
\begin{equation}
 \mathcal{L}^{USCO}(x) = - \ln(1 +x) \; . 
\end{equation}
Note that $U_k (\phi)$ is defined up to an  additive constant, \textit{i. e.} independent of the field $\phi$,
which was discarded from Eq.~\eqref{flow_USCO}.

Since, in the limit $k \to 0 $
\begin{align}
 \mathcal{D}(\epsilon) &  \sim \int_{\mathbf{q}} \delta(k^2 -\mathbf{q}^2 )/\xi \nonumber \\
                                   &  \sim \frac{2 v_D}{\xi} k^{D-2} \; ,
\end{align}
then,  the flow eq.~\eqref{flow_USCO} reduces to
\begin{equation}
\label{flow_USCO_asympt}
 \partial_t U_k = - 2 v_D  k^D  \mathcal{L}^{USCO}(\omega_k)  \; \; (k \to 0) \; ,
\end{equation}
which is the LPA flow equation with an USCO regulator for the continuous (off-lattice)  theory \cite{Ni1,Ni2,Hasen,Caillol1}.
The fixed-points and critical exponents of the lattice and off-lattice versions of the 2 theories are thus identical.

We now discuss the problem of the initial conditions. The USCO regulator $R_k(\mathbf{q}) = Z = \infty$ for
all $k > k_{\mathrm{max}}= \sqrt{4 D/a^2}$. It transpires from the discussion of  sect.~\eqref{LMD}   that the mean-field
solution $U_k(\phi)=U_0(\phi)$  should be solution of the flow-equation~\eqref{flow_USCO} for al
 $\Lambda \geq k \geq k_{\mathrm{max}}$.
Indeed for $k \geq k_{\mathrm{max}}$ we obviously have $\mathcal{D}(\epsilon_k) \equiv 0$ from which
$ \partial_t U_k =0$ follows. Any $\Lambda \geq  k_{\mathrm{max}} $ (with  $R_{\Lambda}= \infty $) can be kept as a valid initial condition
since the MF solution $U_{\Lambda}=U_0$ does not evolves in the range $\Lambda \geq  k  \geq k_{\mathrm{max}} $.

A last remark is in order. The initial condition $U_{k_{\mathrm{max}}}(\phi)= (r/2) \phi^2 + (g/4!) \phi^4$ yields a solution of the flow 
equation~\eqref{flow_USCO} for $k \leq k_{\mathrm{max}}$ only if $1+\omega_{k_{\mathrm{max}}} (\phi) \geq 0$ for all $\phi$.
Therefore the LPA with ultra sharp cut-off is defined for a negative $r$ only if $\vert r \vert <k_{\mathrm{max}}^2 = 4D/a^2 $; otherwise
the flow equation has no solution.

\section{\label{Change}A crucial change of variables}
\begin{figure}[t!]
\includegraphics[angle=0,scale=0.50]{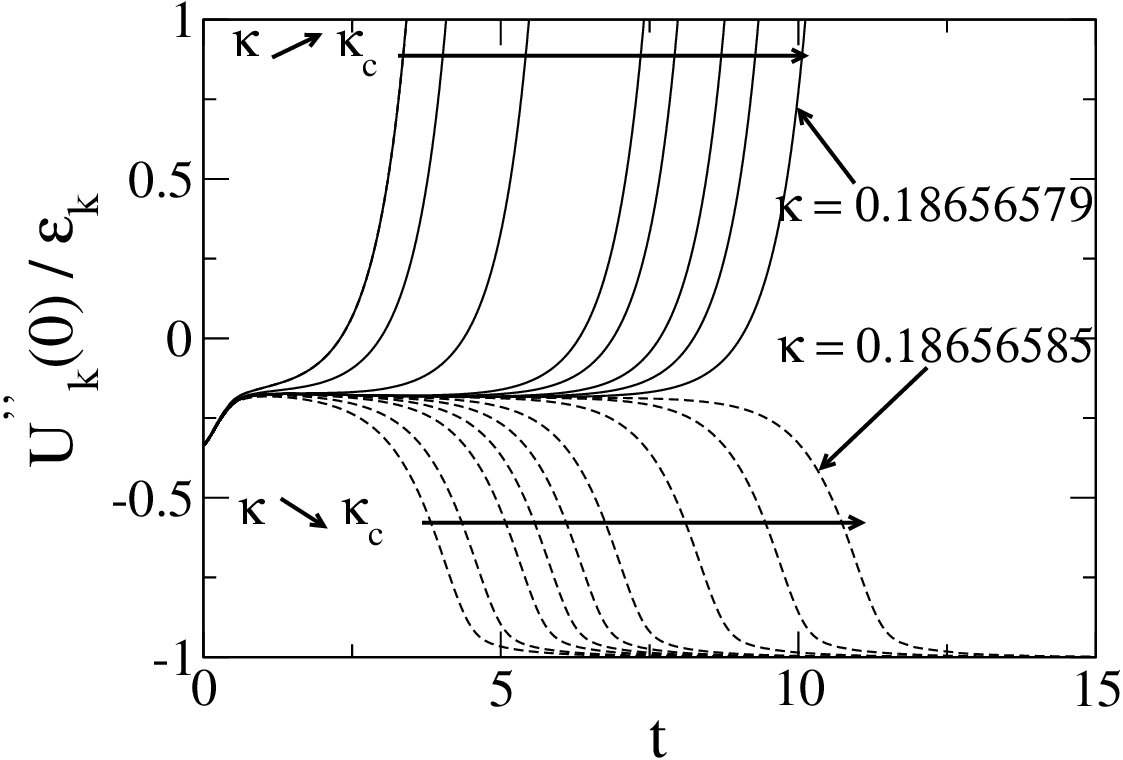}
\caption{
\label{figrc}  $U_k^{''}(M=0)/ \epsilon_k  $ as a function of the RG time ``t'' in the LPA  approximation
with LMD regulator.
$\lambda=1.145$ is fixed and $\kappa_c$ is obtained by dichotomy on $\kappa$. 
Dashed lines : $\kappa > \kappa_c$. Solid lines : $\kappa< \kappa_c$.
}
\end{figure}
We pointed out  in section~\ref{LITIM} and~\ref{USCO} that in the asymptotic limit $k \to 0$
the lattice and off-lattice LPA flow equations bear the same form, both with LMD and USCO regulators.
In the ordered phase this behavior is singular and has been studied at length in Refs~\cite{Bonanno,Caillol1}.
Briefly, in the limit $k \to 0$,  $\omega_k(\phi) = U_k^{''}(\phi)/(\xi k^2) \to -1$ for $-\phi_0(k)<\phi < \phi_0(k) $ where
$\phi_0(k)$ is a precursor of the spontaneous magnetization  $\phi_0 = \lim_{k \to 0}  \phi_0(k)$.
 It follows that the threshold functions $\mathcal{L}$
diverge in this interval as $k^{2-D}$ (for $D>2)$. This yields a universal behavior $\mathcal{L}(\phi)/ \mathcal{L}(\phi=0)
= 1-\phi^2 /\phi_0^2$.  Moreover, as a consequence, $U_{k}(\phi)$ becomes convex as $k \to 0$, in particular
it becomes constant for $-\phi_0 < \phi < + \phi_0$ (see \textit{e. g.} Fig.~\ref{U}).

The divergence of the threshold functions makes impossible to obtain numerical solution of the non-linear 
PDE~\eqref{flow_LMD_f} and \eqref{flow_USCO} in the ordered phase, we really deal with \textit{stiff} equations 
In order to remove stiffness, one is led to make the change of variables 
$U_k(M) \Longrightarrow L_k(M) = \mathcal{L}[   \omega_k(M) \equiv U_k^{''}(M)/\epsilon_k]$. We then obtain the equations
\begin{subequations}
\label{eqs}
\begin{align}
\textrm{(LMD)}   \; \; &  L^{''}_k(\phi) =       \frac{2 \epsilon_k}{\mathcal{N}(\epsilon_k)}    \; \left[ \frac{1}{L_k(\phi) } -1 \right]     +
                                                                \frac{\epsilon_k}{\mathcal{N}(\epsilon_k)} \; \frac{1}{L_k(\phi)^2} \; \partial_t L_k(\phi)
     \; ,  \\
\textrm{(USCO)} \; \; & L^{''}_k(\phi) =         \frac{2 }{\mathcal{D}(\epsilon_k)}  \; \left[  \exp\left(- L_k(\phi) \right)  -1     \right]      +
                                                                   \frac{1 }{\mathcal{D}(\epsilon_k)}  \;  \exp\left(- L_k(\phi) \right)   \; \partial_t L_k(\phi)    \; ,
\end{align}
\end{subequations}
where $k= \Lambda e^{-t}$.

By contradistinction with eqs~\eqref{flow_LMD_final} and~\eqref{flow_USCO} these quasi-linear parabolic PDE can easily be integrated out. 
As in Refs.~\cite{Bonanno,Caillol1}
 we made use of the fully  implicit predictor-corrector algorithm of Douglas-Jones~\cite{Douglas}. This algorithm is unconditionally
stable and convergent and introduces an error of $\mathcal{O}( (\Delta t)^2 )  + \mathcal{O}((\Delta \phi)^2)$
 ($\Delta t$ and $\Delta \phi$ discrete RG time and field steps
respectively) and can be used below and above the critical point as well. 

In order to solve eqs.~\eqref{eqs} numerically one must precise the initial and boundary conditions.
\begin{itemize}
 \item{(i)}\textit{Initial conditions} : they were discussed thoroughly in Sec.~\ref{LPA_Sec}; we have just to transpose this discussion
to the threshold functions. For the LMD regulator one has, at $t=0$,
 $ L_{k_{\mathrm{max}}}(\phi) = \mathcal{L}^{LMD}(U^{\mathrm{loc}\; ''}_{k_{\mathrm{max}}}(\phi) / \epsilon_0^{\mathrm{max}})$ for all 
$\vert \phi \vert \leq \phi_{\mathrm{max}}$,
where $\phi_{\mathrm{max}}$ is the largest value of the field. In practice the local approximation $U_{\mathrm{loc}}(\phi))$ and 
its derivatives with respect to the field
are computed by integrating the exact flow~\eqref{local_implicit} from its MF expression at some large $\Lambda$. 
For the USCO regulator one has, at $t=0$,
$ L_{k_{\mathrm{max}}}(\phi) = \mathcal{L}^{LMD}(U^{''}_{\mathrm{0}}(\phi))  / \epsilon_0^{\mathrm{max}})   $ 
for all $\vert \phi \vert \leq \phi_{\mathrm{max}}$, \textit{i. e.}
one retains the mean field approximation of the potential.
 \item{(ii)} \textit{Boundary conditions} :  for the LMD regulator  we adopted $ L_{k}(\pm  \phi_{\mathrm{max}} ) =
 \mathcal{L}^{LMD}(U^{\mathrm{loc} \; ''}_k(\phi_{\mathrm{max}} ) / \epsilon_k)$ for all $k$ and
for the USCO regulator $ L_{k}(\pm  \phi_{\mathrm{max}} ) =
 \mathcal{L}^{LMD}(U^{''}_{0}(\phi_{\mathrm{max}} ) / \epsilon_k)$ for all $k$ \cite{Caillol1}. It amounts to
keep the first term in the hopping parameter $(\kappa) $ or loop expansions of $\Gamma_k(\phi)$ respectively,
which is a reasonnable assumption at large fields.
\end{itemize}

In order to determine the critical point one proceeds by dichotomy : for instance
$g$ is fixed and one varies $r$. The renormalized coupling constant $\omega_0 =U_k^{''}(M=0)/ \epsilon_k  $ 
in the limit $k \to 0$ discriminates the state : 
for $r>r_c(g)$, $\omega_0 \to \infty$, and for  $r<r_c(g)$, $\omega_0 \to -1$. Alternatively one can fix $\lambda$ 
and vary $\kappa$; an example
is given in fig.~\ref{figrc}. 

\section{\label{Num}Numerical results}
\begin{figure}[t!]
\includegraphics[angle=0,scale=0.55]{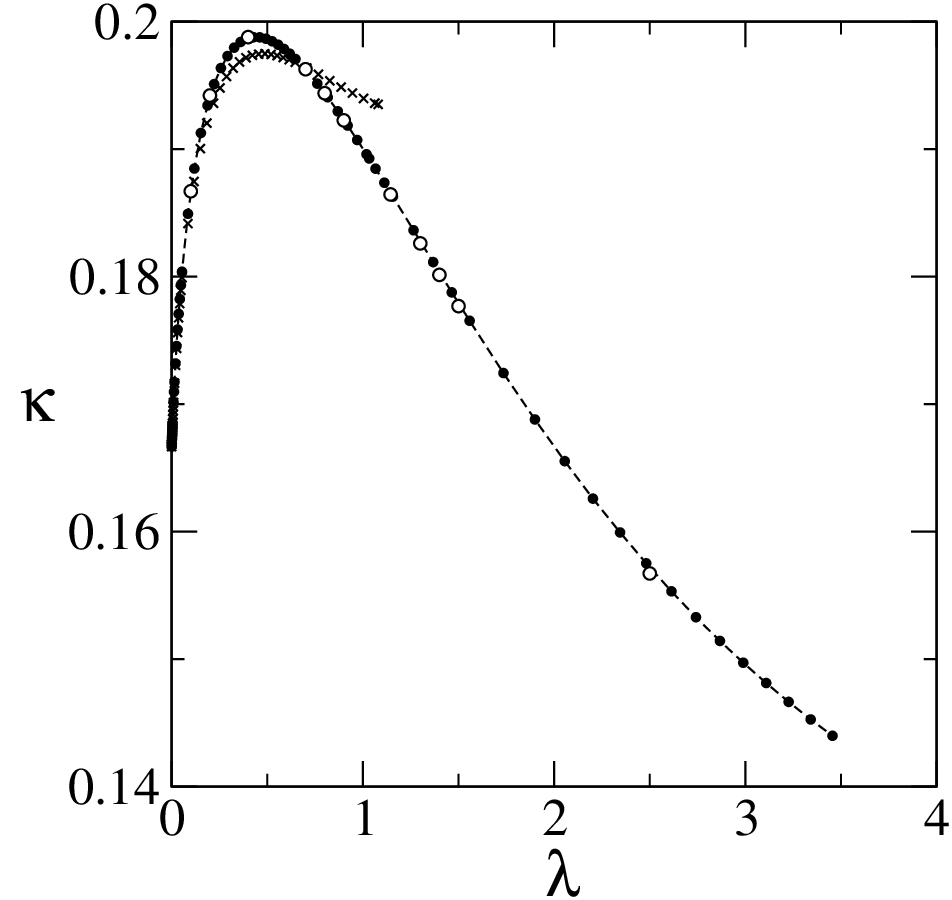}
\caption{\label{cr_line} Critical line $\kappa_c(\lambda)$ of the $\Phi^4$ model on a 3D simple cubic lattice.
Open circles : MC data of Ref.\ \cite{Hasenbusch}; crosses : LPA with USCO regulator; solid circles :
 LPA with LMD regulator, a dashed line joints the points as a guide-line for the eyes.
Uncertainties are smaller than the sizes of the symbols.
}
\end{figure}
We solved Eqs~\eqref{eqs} with the Douglas-Jones algorithm \cite{Douglas} in $D=3$ dimensions of space. To fix the ideas we
used   for most  our numerical experiments    
 $\Delta t = 10^{-4}$, a maximum of $N_t=1.8 10^5$ time steps, $\Delta \phi = 2. 10^{-4}$ and $N_{\phi}=15000$ field steps
(\textit{i. e.} $\phi_{\textrm{max}}=3.$). For the LMD regulator the initial momentum was $\Lambda=5000 a^{-1}$. We checked that
these values of the parameters give at least $7$ stable figures for $\overline{r}_c(g)$. Most our numerical studies were made by
fixing the value of parameter $g$ and varying $r$ in order to determine its critical value $\overline{r}_c(g)$ by dichotomy. 

Our data for the critical line $\overline{r}_c(g)$ are given in table~\ref{Tab_USCO} for the USCO regulator, the eqs.~\eqref{eqs}
has no solution for $\overline{r}_c < -12$ which is a severe drawback. The data for the LMD smooth cut-off are displayed in 
table~\ref{Tab_LMD},  in this case the LPA equations admit solutions for all values of $g$ and we stopped arbitrary our investigations
at $g=1000$.
All these data are also displayed with the variables $(\lambda, \kappa) $ in Fig.~\ref{cr_line}  in order to be compared with the MC 
data that  Hasenbusch obtained for several points \cite{Hasenbusch}.
As apparent in Fig.~\ref{cr_line} the theoretical predictions of the LPA  with an USCO regulator (crosses) are poor as soon as 
$\lambda \geq 0.5$. 
We interpret this failure as a consequence of the use of a mean-field initial condition at $\Lambda=k_{\mathrm{max}}$ which turns out to
be a bad approximation of the local theory at high values of $\lambda$.
By contrast a very good overall agreement between the Monte Carlo (MC) Data and the predictions of the LPA with LMD regulator is observed.
This confirm the conclusions of Machado and Dupuis in Ref.~\cite{Dupuis-M} who obtained also such a good agreement in the 
case of the 3D Ising, XY and Heisenberg models. 

 A more stringent comparison is made in  table~\ref{Tab_compa} where, for all the $\lambda$ considered in Ref.~\cite{Hasenbusch}, 
the critical $\kappa_c(\lambda)$ was obtained by dichotomy on $\kappa$. The maximum relative error
of the LPA-LMD theory can be seen not to exceed $\sim 3. 10^{-3}$ for the considered range of parameters.
 In the case of Ising, XY and Heisenberg models Machado and Dupuis report 
errors which are  significantly higher, \textit{i. e.} of the order of a few per cents, than the ones we obtained for the $\Phi^4$ model.
However these authors  used a standard explicit Euler integration scheme
for the non-linear PDE for $U_k$ which yields to stop the flow before its scaling limit $k \to 0$ in the ordered phase \cite{Dupuis-M}.
We suggest that  solving instead the quasi-linear parabolic equations satisfied by  the threshold
functions $L_k$ could perhaps  change the evaluation of the critical parameters, yielding a still better status for the LPA predictions.
This point should be checked.


\section{\label{Conclu} Conclusion}

In this paper we have computed the critical line of the $\Phi^4$ one-component model on the simple  cubic lattice in three dimensions
in the framework of the NPRG within the LPA approximation. We have considered both a sharp and a smooth regulator.
The flow equations have been solved for the threshold functions rather than for the potential. This trick allows to obtain
numerical solutions in the ordered phase where the PDE for the potential are stiff and  fail to converge.  

A dichotomy process based on the generically different
asymptotic  behaviors of the adimensionned susceptibility $U_k^{''}(\phi=0)/k^2$ in zero field, below and above the critical point,
 yields a very precise determination of the (non-universal) critical parameters.

The LPA with a sharp cut-off regulator must be supplemented with mean-field initial conditions at $\Lambda=k_{\mathrm{max}}$ which restricts the solution of the equation 
 to a small domain of the $(r,g)$ plane. Moreover, even in this restricted domain,  the critical parameters are in poor
 agreement with the MC data of Ref.~\cite{Hasenbusch}. When the smooth LMD regulator is considered, the PDE must
be supplemented either with  mean-field initial conditions at $\Lambda=\infty$ or by the exact local expression of the potential at any
$\Lambda \geq k_{\mathrm{max}}$. The various possible initial local conditions obey an exact flow equation which coincides
with the LPA-LMD theory.
The LPA with LMD cut-off gives surprisingly good estimates of the critical parameters of the lattice $\Phi^4$ model,
 the maximum deviation with MC data being of $\sim 3. 10^{-3}$ for the states which we considered.

Extension of the present study to several other  lattices  seem feasible as well as the extension to vectorial $O(N)$ models.

\begin{acknowledgments}
The author  thanks  K.~ Binder and A.~Tr\"{o}ster for  e-mail correspondence. 
\end{acknowledgments}

\appendix*
\section{\label{app1}Exact flow equation for the local theory}
We noted in the text that in the range $\Lambda \geq k \geq k_{\mathrm{max}}$ the average effective action is local
for the LMD regulator.
The local potential $U_k(\phi)$ satisfies exactly to Eq.~\eqref{local_implicit} and  we rewrite this relation as
\begin{align}
 \exp\left( 
-U_k(\phi) + U_k^{'}(\phi) \phi +\frac{1}{2} \epsilon_k \phi^2 
\right)   &=  \int_{-\infty}^{+\infty} d\varphi \; \exp \left( -U_0(\varphi)  -\frac{1}{2} \epsilon_k \varphi^2 +
( U_k^{'}(\phi)+\epsilon_k \phi ) \varphi
\right)  \nonumber \\
&=z_k(h)  \; ,
\end{align}
where $z_k(h)$ is the one-site partition function~\eqref{z} and $h=U_k^{'}(\phi)+ \epsilon_k \phi$ an effective magnetic field.
Taking the derivatives of both sides of this equality  with respect to scale ''$k$'' yields, after rearrangement
\begin{equation}
\label{lo}
 -\partial_k U_k(\phi) + \phi \partial_k U_k^{'}(\phi) + \dfrac{1}{2} \partial_k \epsilon_k \phi^2 =
-\frac{1}{2}\partial_k \epsilon_k <\varphi^2> + (\partial_k U_k^{'}(\phi)  + \phi \partial_k \epsilon_k ) <\varphi>
\end{equation}

Note that we have
\begin{align}
 <\varphi> & = \phi = \partial \ln z/\partial h \nonumber \\
<\varphi^2> - <\varphi>^2 &= \partial \phi/ \partial h  = 1/ ( \partial h /\partial \phi)              \nonumber \\
                                        &= \dfrac{1}{U_k^{''} (\phi) + \epsilon_k} \; .
\end{align}
Inserting these identities into~\eqref{lo} yields 
\begin{equation}
 \partial_k U_k(\phi) = \frac{1}{2} \dfrac{\partial_k \epsilon_k}{U_k^{''}  (\phi)+ \epsilon_k} \; .
\end{equation}
A further simplification occurs with the choice $\epsilon_k \propto k^2$ which entails
\begin{equation}
 \partial_t U_k(\phi) =  -\dfrac{ \epsilon_k}{U_k^{''} (\phi) + \epsilon_k} \; .
\end{equation}

\newpage
\begin{table*}[h!]
\caption{ \label{Tab_USCO}Critical parameters  of the $\Phi^4$ theory on a 3D simple cubic lattice in the LPA approximation using an USCO cut-off.
From left to right : $\overline{g}$, $\overline{r}_c(\overline{g})$, $\kappa_c$, and $\lambda_c$. The data were obtained by fixing $\overline{g}$
and determining  $\overline{r}_c(\overline{g})$ by dichotomy.
$\kappa_c$ and $\lambda_c$
were then obtained from ($\overline{g}$, $\overline{r}_c(\overline{g})$) via Eqs. \eqref{toto}. An uncertainty of $\pm 1$ affects the last figure.}
\begin{tabular}{|l|l|l|l||l|l|l|l|} 
\hline 
  $ \overline{g}$    &      $ \overline{r}_c$          &     $ \kappa_c$                  &   $\lambda_c$   &    
   $ \overline{g}$    &      $\overline{r}_c $          &     $ \kappa_c$                  &   $\lambda_c$          \\ \hline
0.000000        &   0.0                &  0.166667   &  0.0                                      &   0.200000 $10^2$  &  -0.191553 $10^1$    &  0.187467   &  0.117147           \\  \hline
0.100000        &  -0.125538 $10^{-1}$ &  0.166861   &  0.464044 $10^{-3}$ &   0.250000 $10^2$  &  -0.232170 $10^1$    &  0.190042   &  0.150484      \\  \hline
0.200000        &  -0.249786 $10^{-1}$ &  0.167052   &  0.930213 $10^{-3}$ &  0.300000 $10^2$  &  -0.271343 $10^1$    &  0.192047   &  0.184411     \\  \hline
0.300000        &  -0.372955 $10^{-1}$ &  0.167240   &  0.139846 $10^{-2}$ &  0.350000 $10^2$  &  -0.309339 $10^1$    &  0.193600   &  0.218640     \\ \hline
0.400000        &  -0.495161 $10^{-1}$ &  0.167425   &  0.186875 $10^{-2}$ &  0.400000 $10^2$  &  -0.346354 $10^1$    &  0.194792   &  0.252959     \\  \hline
0.500000        &  -0.616484 $10^{-1}$ &  0.167608   &  0.234105 $10^{-2}$ &  0.450000 $10^2$  &  -0.382536 $10^1$    &  0.195693   &  0.287219   \\  \hline
0.600000        &  -0.736986 $10^{-1}$ &  0.167789   &  0.281532 $10^{-2}$ &  0.500000 $10^2$  &  -0.418000 $10^1$    &  0.196361   &  0.321312  \\  \hline
0.700000        &  -0.856719 $10^{-1}$ &  0.167968   &  0.329154 $10^{-2}$ &  0.550000 $10^2$  &  -0.452842 $10^1$    &  0.196839   &  0.355168  \\  \hline
0.800000        &  -0.975725 $10^{-1}$ &  0.168144   &  0.376968 $10^{-2}$  &  0.600000 $10^2$  &  -0.487136 $10^1$    &  0.197164   &  0.388737    \\  \hline
0.900000        &  -0.109404           &  0.168319   &  0.424970 $10^{-2}$        &  0.650000 $10^2$  &  -0.520949 $10^1$    &  0.197365   &  0.421990     \\  \hline
0.100000 $10^1$  &  -0.121170           &  0.168492   &  0.473160 $10^{-2}$   & 0.700000 $10^2$  &  -0.554300 $10^1$    &  0.197458   &  0.454881  \\  \hline
0.120000 $10^1$  &  -0.144516           &  0.168833   &  0.570091 $10^{-2}$   & 0.750000 $10^2$  &  -0.587340 $10^1$    &  0.197484   &  0.487499    \\  \hline
0.140000 $10^1$  &  -0.167631           &  0.169167   &  0.667742 $10^{-2}$     &   0.800000 $10^2$  &  -0.620008 $10^1$    &  0.197437   &  0.519751      \\  \hline
0.160000 $10^1$  &  -0.190530           &  0.169495   &  0.766098 $10^{-2}$       &  0.850000 $10^2$  &  -0.652375 $10^1$    &  0.197337   &  0.551677    \\  \hline
0.180000 $10^1$  &  -0.213227           &  0.169818   &  0.865143 $10^{-2}$       &   0.900000 $10^2$  &  -0.684407 $10^1$    &  0.197183   &  0.583218     \\  \hline
0.200000 $10^1$  &  -0.235735           &  0.170135   &  0.964863 $10^{-2}$       &  0.950000 $10^2$  &  -0.716333 $10^1$    &  0.197020   &  0.614599   \\  \hline
0.300000 $10^1$  &  -0.345763           &  0.171648   &  0.147315 $10^{-1}$      &  0.100000 $10^3$  &  -0.747982 $10^1$    &  0.196819   &  0.645628     \\  \hline
0.400000 $10^1$  &  -0.452243           &  0.173055   &  0.199654 $10^{-1}$       & 0.110000 $10^3$  &  -0.810742 $10^1$    &  0.196364   &  0.706911    \\  \hline
0.500000 $10^1$  &  -0.555797           &  0.174373   &  0.253384 $10^{-1}$       & 0.120000 $10^3$  &  -0.872935 $10^1$    &  0.195870   &  0.767298  \\  \hline
0.600000 $10^1$  &  -0.656858           &  0.175612   &  0.308396 $10^{-1}$       &  0.130000 $10^3$  &  -0.934719 $10^1$    &  0.195365   &  0.826962    \\  \hline
0.700000 $10^1$  &  -0.755749           &  0.176780   &  0.364598 $10^{-1}$       & 0.140000 $10^3$  &  -0.996241 $10^1$    &  0.194871   &  0.886080   \\  \hline 
0.800000 $10^1$  &  -0.852719           &  0.177884   &  0.421903 $10^{-1}$       & 0.150000 $10^3$  &  -0.105764 $10^2$    &  0.194406   &  0.944839             \\  \hline
0.900000 $10^1$  &  -0.947970           &  0.178929   &  0.480232 $10^{-1}$       & 0.160000 $10^3$  &  -0.111905 $10^2$    &  0.193981   &  0.100343 $10^1$  \\  \hline
0.100000 $10^2$  &  -0.104167 $10^1$    &  0.179919   &  0.539513 $10^{-1}$  & 0.170000 $10^3$  &  -0.118061 $10^2$    &  0.193609   &  0.106206 $10^1$ \\  \hline
0.150000 $10^2$  &  -0.149095 $10^1$    &  0.184166   &  0.847930 $10^{-1}$  & 0.172979 $10^3$  &  -0.119900 $10^2$    &  0.193509   &  0.107956 $10^1$  \\  \hline
\end{tabular}
\end{table*}

\newpage
\begin{table*}[h!]
 \caption{ \label{Tab_LMD}Critical parameters  of the $\Phi^4$ theory on a 3D simple cubic lattice in the LPA approximation using an LMD cut-off.
From left to right : $\overline{g}$, $\overline{r}_c(\overline{g})$, $\kappa_c$, and $\lambda_c$. The data were obtained by fixing $\overline{g}$
 and determining $\overline{r}_c(\overline{g})$ by dichotomy.
$\kappa_c$ and $\lambda_c$
were then obtained from ($\overline{g}$, $\overline{r}_c( \overline{g}) $) via Eqs. \eqref{toto}. An uncertainty of $\pm 1$ affects the last figure.}
\begin{tabular}{|l|l|l|l||l|l|l|l|} 
 \hline 
  $ \overline{g} $    &      $ \overline{r}_c$          &     $ \kappa_c$                  &   $\lambda_c$     &       
     $\overline{g}  $    &      $\overline{r}   $          &     $ \kappa_c$     &   $\lambda_c$          \\ \hline
0.000000         &    0.0                            &   0.166667   &   0.0                           &   0.800000 $10^2$  &   -0.625341 $10^1$     &   0.198459   &   0.525147                 \\ \hline
0.100000         &   -0.125603 $10^{-1}$  &   0.166861   &   0.464045 $10^{-3}$ &   0.850000 $10^2$  &   -0.656974 $10^1$     &   0.198190   &   0.556459   \\ \hline
0.200000         &   -0.250055 $10^{-1}$  &   0.167053   &   0.930221 $10^{-3}$ &   0.900000 $10^2$  &   -0.688192 $10^1$     &   0.197864   &   0.587250    \\ \hline
0.300000         &   -0.373541 $10^{-1}$  &   0.167242   &   0.139849 $10^{-2}$  &   0.950000 $10^2$  &   -0.719022 $10^1$     &   0.197488  &   0.617527   \\ \hline
0.400000         &   -0.496161 $10^{-1}$  &   0.167428   &   0.186881 $10^{-2}$ &   0.100000 $10^3$  &   -0.749487 $10^1$     &   0.197074   &   0.647300  \\ \hline
0.500000         &   -0.617986 $10^{-1}$  &   0.167613   &   0.234117 $10^{-2}$ &   0.110000 $10^3$  &   -0.809410 $10^1$     &   0.196151   &   0.705380   \\ \hline
0.600000         &   -0.739072 $10^{-1}$  &   0.167795   &   0.281552 $10^{-2}$ &   0.120000 $10^3$  &   -0.868108 $10^1$     &   0.195140   &   0.761593   \\ \hline
0.700000         &   -0.859458 $10^{-1}$  &   0.167976   &   0.329184 $10^{-2}$ &   0.130000 $10^3$  &   -0.925704 $10^1$     &   0.194072   &   0.816050   \\ \hline
0.800000         &   -0.979189 $10^{-1}$  &   0.168154   &   0.377011 $10^{-2}$ &   0.140000 $10^3$  &   -0.982301 $10^1$     &   0.192969   &   0.868861   \\ \hline
0.900000         &   -0.109830                 &   0.168331    &   0.425031 $10^{-2}$ &    0.150000 $10^3$  &  -0.103800 $10^2$    &    0.191849 &     0.920148 \\ \hline
0.100000 $10^1$  &   -0.121680            &   0.168507   &   0.473241 $10^{-2}$ &    0.160000 $10^3$  &   -0.109284 $10^2$     &   0.190719   &   0.969966  \\ \hline
0.200000 $10^1$  &   -0.237346            &   0.170181   &   0.965382 $10^{-2}$ &   0.170000 $10^3$  &   -0.114700 $10^2$      &   0.189603   &   0.101856 $10^1$  \\ \hline
0.250000 $10^1$  &   -0.293536            &   0.170971   &   0.121796 $10^{-1}$  &   0.172979 $10^3$  &   -0.116300 $10^2$     &   0.189272   &   0.103280 $10^1$ \\ \hline
0.300000 $10^1$  &   -0.348802            &   0.171735   &   0.147464 $10^{-1}$  &   0.180000 $10^3$  &   -0.120028 $10^2$     &   0.188475   &   0.106569 $10^1$  \\ \hline
0.400000 $10^1$  &   -0.456919            &   0.173190   &   0.199965 $10^{-1}$  &   0.190000 $10^3$  &   -0.125300 $10^2$     &   0.187373   &   0.111177 $10^1$ \\ \hline
0.500000 $10^1$  &   -0.562248            &   0.174560   &   0.253927 $10^{-1}$  &  0.200000 $10^3$  &   -0.130506 $10^2$      &   0.186282   &   0.115670 $10^1$ \\ \hline
0.600000 $10^1$  &   -0.665180            &   0.175854   &   0.309247 $10^{-1}$ &   0.225000 $10^3$  &   -0.143280 $10^2$      &   0.183644   &   0.126469 $10^1$  \\ \hline
0.700000 $10^1$  &   -0.766006            &   0.177080   &   0.365834 $10^{-1}$ &   0.250000 $10^3$  &   -0.155743 $10^2$      &   0.181139   &   0.136714 $10^1$   \\ \hline
0.800000 $10^1$  &   -0.865000            &   0.178243   &   0.423608 $10^{-1}$ &   0.275000 $10^3$  &   -0.167936 $10^2$      &   0.178772   &   0.146480 $10^1$  \\ \hline
0.900000 $10^1$  &   -0.962205            &   0.179345   &   0.482472 $10^{-1}$ &   0.300000 $10^3$  &  -0.179892 $10^2$       &  0.176537    &   0.155827 $10^1$   \\ \hline
0.100000 $10^2$  &   -0.105791 $10^1$     &   0.180395   &   0.542370 $10^{-1}$  &   0.350000 $10^3$ &  -0.203196 $10^2$  &  0.172445    &  0.173467 $10^1$ \\ \hline
0.150000 $10^2$  &   -0.151715 $10^1$     &   0.184929   &   0.854965 $10^{-1}$ &   0.400000 $10^3$  &  -0.225821 $10^2$  &  0.168797    &  0.189951 $10^1$   \\ \hline
0.200000 $10^2$  &   -0.195105 $10^1$     &   0.188483   &   0.118420    &     0.450000 $10^3$  &   -0.247888 $10^2$     &   0.165533   &   0.205508 $10^1$  \ \\ \hline
0.250000 $10^2$  &   -0.236560 $10^1$     &   0.191267   &   0.152429    &     0.500000 $10^3$  &   -0.269479 $10^2$     &   0.162590   &   0.220296 $10^1$   \\ \hline
0.300000 $10^2$  &   -0.276459 $10^1$     &   0.193433   &   0.187082    &     0.550000 $10^3$  &   -0.290700 $10^2$     &   0.159940   &   0.234491 $10^1$  \\ \hline
0.350000 $10^2$  &   -0.315063 $10^1$     &   0.195101   &   0.222043    &     0.600000 $10^3$  &   -0.311562 $10^2$     &   0.157522   &   0.248133 $10^1$   \\ \hline
0.400000 $10^2$  &   -0.352561 $10^1$     &   0.196364   &   0.257059    &     0.650000 $10^3$  &   -0.332136 $10^2$     &   0.155317   &   0.261337 $10^1$  \\ \hline
0.450000 $10^2$  &   -0.389103 $10^1$     &   0.197299   &   0.291951    &     0.700000 $10^3$  &   -0.352456 $10^2$     &   0.153295   &   0.274161 $10^1$ \\ \hline
0.500000 $10^2$  &   -0.424800 $10^1$     &   0.197965   &   0.326583    &     0.750000 $10^3$  &   -0.372553 $10^2$     &   0.151435   &   0.286658 $10^1$ \\ \hline
0.550000 $10^2$  &   -0.459733 $10^1$     &   0.198407   &   0.360850    &     0.800000 $10^3$  &   -0.392455 $10^2$     &   0.149718   &   0.298872 $10^1$  \\ \hline
0.600000 $10^2$  &   -0.494000 $10^1$     &   0.198672   &   0.394704    &     0.850000 $10^3$  &   -0.412184 $10^2$     &   0.148127   &   0.310840 $10^1$ \\ \hline
0.650000 $10^2$  &   -0.527631 $10^1$     &   0.198782   &   0.428072    &     0.900000 $10^3$  &   -0.431761 $10^2$     &   0.146650   &   0.322593 $10^1$  \\ \hline
0.700000 $10^2$  &   -0.560707 $10^1$     &   0.198771   &   0.460949    &     0.950000 $10^3$  &   -0.451203 $10^2$     &   0.145275   &   0.334160 $10^1$  \\ \hline
0.750000 $10^2$  &   -0.593264 $10^1$     &   0.198657   &   0.493310    &     0.100000 $10^4$  &   -0.470524 $10^2$     &   0.143992   &   0.345561 $10^1$  \\ \hline
\end{tabular}
\end{table*}

\newpage
\begin{table*}[h!]
\caption{\label{Tab_compa}Critical line  $ \kappa_c(\lambda)$  of the $\Phi^4$ model on a 3D simple cubic lattice.
The MC data for $\kappa_c$ (second column) are those of Ref.\ \cite{Hasenbusch}.
The LPA data  reported in the third column were obtained with the use of LMD regulator by fixing $\lambda$ and 
determining $\kappa_c(\lambda)$ by dichotomy.
The (signed) relative error is defined as 
$\epsilon = ( \kappa_{c, \textrm{LPA}} - \kappa_{c, \textrm{MC}})/\kappa_{c, \textrm{MC}}$.}
\begin{tabular}{|l|l|l|l|} 
\hline 
 $ \lambda $    &      $ \kappa_ {c, \textrm{MC}}$      &     $ \kappa_ {c, \textrm{LPA}}$               &   $\epsilon$                  \\ \hline
0.1       &    0.18670475      &      0.1866196     &    -0.000456    \\ \hline
0.2       &    0.19421255      &      0.1941031     &    -0.000564    \\ \hline
0.4       &    0.19879185      &      0.1986966     &    -0.000479    \\ \hline
0.7       &    0.19626510      &      0.1962421     &    -0.000117    \\ \hline
0.8       &    0.19438785      &      0.1943946     &     0.000003    \\ \hline
0.9       &    0.19225565      &      0.1922928     &     0.000193    \\ \hline
1.145   &    0.18644630      &      0.1865659     &     0.000641    \\ \hline
1.3       &    0.18261165      &      0.1827799     &     0.000921    \\ \hline
1.4       &    0.18013945      &      0.1803395     &     0.001111    \\ \hline
1.5       &    0.17769270      &      0.1779256     &     0.001311    \\ \hline
2.5       &    0.15671735      &      0.1572028     &    0.003098     \\ \hline
\end{tabular}
\end{table*}

\newpage


\end{document}